\documentstyle[12pt,epsf]{article}
\begin{document}
\textwidth 15 true cm
\textheight 23 true cm
\baselineskip=10pt
\voffset=-30 mm
\renewcommand{\baselinestretch}{1.0}
\renewcommand{\theequation}{\arabic{equation}}
\def\brho{\vec{\mbox{\boldmath $\rho $}}}
\def\br{\vec{\mbox{\boldmath $r$}}}
\def\bn{\vec{\mbox{\boldmath $n$}}}
\def\bQ{\vec{\mbox{\boldmath $Q$}}}
\def\imathr{\dot{\imath}}

\title{\bf{Isotropic - Nematic Transition of Surface Embedded
Polymers}
\\{ and the Associated Tubulization Transition of the Embedding Surface}}
\author{ R.Podgornik ~\thanks{E-mail address: rudolf.podgornik@ijs.si}
\\{Department of Theoretical Physics}
\\{J. Stefan Institute}
\\{1000 Ljubljana, Slovenia}}
\begin{titlepage}
\maketitle
\baselineskip=8pt
\renewcommand{\baselinestretch}{1.0}
\begin{abstract}
\small
A self-interacting polymer can undergo an orientational ordering
transition, depending on the magnitude of the nematic interaction. The
effect of embedding such a polymer into a flexible surface on this
transition is studied on the mean-field level. Renormalized values of
the elastic constants of the ``dressed'' surface are derived as
functions of the orientational order parameter of the polymer
chain. In the disordered state the surface tension and curvature
modulus remain scalars but depend on the surface coverage of the
embedded polymer. In the nematic state there is a symmetry breaking
transition leading to anisotropic elastic constants. At a sufficiently
large nematic order parameter the effective surface tension in the
direction perpendicular to the nematic axis can become negative,
leading to tubulization of the embedding surface.  
\end{abstract}
\end{titlepage}
\textwidth 17 true cm
\textheight 23 true cm
\baselineskip=15pt
\parskip=20pt plus 1pt 
\parindent=10pt

\section{Introduction}

Usually the adsorption of polymers onto surfaces is discussed in terms
of more or less diffuse adsorption layers, composed of trains, loops
and dangling ends. There is enough corroborating evidence that in many
cases this simplified picture correctly captures the salient features
of the polymer adsorption phenomena \cite{ref0}.

However, some recent results on adsorption of highly charged
polyelectrolyes onto oppositely charged molecular surfaces give
evidence to much stronger bonding of the adsorbing polymer to the
surface \cite{ref2,ref3}. It seems that the polymer adsorbs flat onto
the surface without any observable loops, dangling ends and even a
very restricted number of crossings among the different segments of
the chain, as the strong electrostatic attraction can apparently
override any entropy driven tendency for less tightly bound
configurations.

In this respect one can envision the adsorption process more as an
embedding of polymer chains into (or onto) the adsorbing (supporting)
surface. The theoretical significance of this observation would be
that an analysis of the adsorption phenomenon could be carried out
without taking into consideration the extremely improbable
configurations with local desorption of the polymers (loops and
dangling ends). In certain sense this simplifies the analysis because
the chains can be treated as embedded into the supporting
matrix. However, the embedding also signifies that there is a strong
coupling between the polymer and the surface configurations and thus
the pertaining degrees of freedom can not be decoupled and treated
separately.

In this contribution we shall address the interplay of the polymer and
surface degrees of freedom in connection with the nematic ordering
transition of a polymer chain embedded ({\sl i.e.} tightly adsorbed)
onto a flexible, fluid surface. It will be established that there can
be no effective decoupling of the tight adsorption problem (embedding)
into a separate polymer and surface parts. The influence of the
surface degrees of freedom on the polymer ordering transition, as well
as of the polymer degrees of freedom on the effective elastic
constants of the ``dressed'' (meaning surface + polymer) surface will
be derived and discussed. The renormalized form of the surface elastic
constants will be shown to lead to a tubulization transition of the
embedding surface.

\section{Summary}

I first present a sketchy overview of the main results and conclusions
reached in the paper. Starting from an orientationally (nematically)
interacting polymer chain embedded into a soft (membrane) surface,
conditions for a second order symmetry breaking orientational ordering
transition are derived. The presence of surface shape fluctuations
only marginally influences the critical point of the transition. The
nematic orientational order parameter, defined in a standard way
\cite{prost} as $
S = \frac{1}{2}\left( 3 \mathopen< \cos^2{\Theta}\mathclose>- 1
\right)$, where $\Theta$ is the angle between the local direction and
the nematic axis, is obtained as a solution of 
\begin{equation}
S = \sqrt{1 - \sqrt{\frac{1}{2(\beta\epsilon\bar{\rho}) \beta u'_a}}},
\end{equation}
where $\epsilon$ is the elastic modulus of the polymer chain,
$\bar{\rho}$ is the polymer surface density (surface coverage) and
$u'_a$ is proportional to the excluded volume of the nematic
interaction $u_a$, renormalized due to the presence of the embedding surface
shape fluctuations,
\begin{equation}
\beta u'_a \longrightarrow \beta u_a -
{\textstyle\frac{1}{2}}\frac{\beta\epsilon}{\bar{\rho}} \mathopen<
\left(  c_i - c_k\right)^2 \mathclose>_0.
\label{urenor}
\end{equation}
where $c_i$ is the embedding surface curvature in the direction $i$, where $
i = x,y$. The average in the above equation is with respect to the
bare surface, {\sl i.e.} with no embedded polymers.

Tied to this transition are also changes in the overall (``dressed'')
elastic constants of the membrane. A closed expression is derived for
the renormalized elastic constants, surface tension ($\gamma$) and the
elastic modulus ($K_c$), of the ``dressed'' membrane
\begin{eqnarray}
\gamma &\longrightarrow& \gamma + 2\frac{\lambda}{\beta} \bar{\sigma}_{\alpha} +
\epsilon \bar{c}_{\alpha} \nonumber\\ 
K_c &\longrightarrow& K_c +
\frac{\epsilon}{\bar{\rho}}\bar{\sigma}_{\alpha}\bar{\sigma}_{\beta},
\end{eqnarray}
where the index $\alpha$ describes the components of the indexed
quantities in the reference frame of their eigen-coordinate
system. Thus $\alpha$ can be either $n$ signifying a component
parallel to the nematic axis, or $\perp$ signifying a component
perpendicular to the nematic axis. $\bar{\sigma}_{\alpha}$ is the
mean-field orientational tensor of the embedded polymer, and
$\bar{c}_{\alpha}$ its mean-field squared curvature tensor. $\lambda$
is a scalar depending on the orientational order parameter.

Following from the above expression for the renormalized values of the
elastic constants, we show that the renormalized surface tension can
change sign in the direction perpendicular to the nematic axis for a
sufficiently large nematic order parameter. Basing our analysis on the
recent general investigation of the tubulization transition of fluid
membranes by Radzihovsky and Toner \cite{ref7} we show,
in the framework of a Landau theory, that the preferred conformation of
the surface is a tubule with the long axis perpendicular to the
nematic direction of the embedded polymers. The tubulization
transition of the membrane does not coincide with the isotropic -
nematic transition of the embedded polymers.

\section{Analysis}

\subsection{Outline}

The formalities of the derivation are a bit convoluted so I will give
a short guided tour before actually developing them. The point of
departure is the usual flexible chain Hamiltonian together with the
``embedding {\sl ansatz}'', constraining the chain to lie on the
supporting surface. The expression for the partition function
following from these premises is developed in Section 3.2. A closed
form of the partition function can be obtained by introducing
collective variables of the polymer chain: density, orientational
tensor and tensor of the squared curvature. This is developed in
Section 3.3 together with the final form of the partition function in
collective coordinates in Section 3.4. This partition function has a
simple closed form solution if the collective variables are treated in
a mean-field manner, ignoring local fluctuations in their values along
the surface. This is done in Section 3.5 leading to the final
derivation of effective elastic constants of the ``dressed'' membrane
in Section 3.6. Once these are derived the tubulization transition
can be described by means of the formalism set forth by Radzihovsky and
Toner \cite{ref7} as is done in Section 3.7.

\subsection{Theory and Formalism}

The position of the $n-th$ bead along the polymer contour is described
by a position vector $\br(n) = (x(n),y(n),z(n))$, which is a
continuous, differentiable function of the arclength $n$, while the
total length of the polymer is $N$. I limit myself to a single polymer
chain case, but a generalization to many chains is quite
straightforward. Let the embedding surface be represented in the Monge
parametrization as $ z = \zeta (x,y) = \zeta (\brho) $, where $\brho$
is the two-dimensional radius vector. In this notation the partition
function of a flexible, but otherwise noninteracting, polymer chain
embedded into the surface $\zeta (\brho)$ can be written as \cite{ref1}
\begin{equation}
\Xi\left(\zeta \left(\brho (n)\right)\right) = \int\!\! \dots \!\!\int {\cal
D}{\br}(n) \prod_{n}\delta\left(z(n) - \zeta(\brho (n))\right)
\prod_{n}\delta\left( \dot{\br}(n)^2 - 1 \right) \times
e^{-\beta{\cal H}_0({\br}(n))}.
\label{equ1}
\end{equation}
where
\begin{equation}
\beta{\cal H}_0({\br}(n)) = {\textstyle \frac{1}{2}}\beta\epsilon
\int_{0}^{N} \left(\frac{d^2{\br}(n)}{dn^2}\right)^2~dn =
{\textstyle\frac{1}{2}}\beta\epsilon \int_{0}^{N}
\left(\ddot{\br}(n)\right)^2~dn.
\label{equ2}
\end{equation}
$\beta \epsilon$ is the energy associated with polymer bending and can
be related to the persistence length \cite{ref1}. By integrating out
the $z(n)$ variable and enforcing the continuity condition
$\dot{\br}(n)^2 = 1$ only globally \cite{ref6}, {\sl i.e.}
\begin{equation}
\left< \dot{\br}(n)^2 \right> = \left< \dot{\brho}(n)^2 +
\dot{\zeta}\left(\brho(n)\right)^2 \right> = 1,
\label{equ3}
\end{equation}
one is led to the following form of the corresponding partition function,
\begin{equation}
\Xi\left(\zeta \left(\brho\right)\right) = \int\!\! \dots
\!\!\int{\cal D}\brho(n)~exp\left(- \beta{\cal H}_0(\brho (n)) \right),
\label{equ4}
\end{equation}
where some irrelevant constant terms have been omitted. The effective
Hamiltonian ${\cal H}_0(\brho (n))$ has been obtained \cite{ref1} as
\begin{eqnarray}
\beta{\cal H}_0(\brho (n)) &=&
{\textstyle\frac{1}{2}}\beta\epsilon \int_{0}^{N} \left(
\ddot{\brho}(n)^2 + \ddot{\zeta}(n)^2 \right)dn~ + ~\lambda\!\!\int_{0}^{N}\!\!  \left(\dot{\brho}(n)^2 + \dot{\zeta}(n)^2 \right)dn~  \nonumber\\
&=& {\textstyle\frac{1}{2}}\beta\epsilon \int_{0}^{N} \left(
\ddot{\brho}(n)^2 + \left(\frac{\partial \zeta(\brho )}{\partial
\brho_i}~\ddot{\brho}_i(n) +  \frac{\partial^2\zeta(\brho )}{\partial \brho_i~\partial \brho_k}~\dot{\brho}_i(n)\dot{\brho}_k(n) \right)^2\right)dn~ + \nonumber\\
&+& ~\lambda\!\!\int_{0}^{N}\!\!  \left(\dot{\brho}(n)^2 + \frac{\partial
\zeta(\brho )}{\partial 
\brho_i}\frac{\partial \zeta(\brho )}{\partial \brho_k}
\dot{\brho}_i(n)\dot{\brho}_k(n) \right)dn~, \nonumber\\
~ 
\label{equ5}
\end{eqnarray}
where the constant $\lambda$ is the Lagrange multiplyer, assuring the
global condition Eq. \ref{equ3}, to be determined later. The validity
of the above Hamiltonian is restricted to the case of a global
continuity constraint Eq. \ref{equ3} (for details of this
transformation see \cite{ref1} and \cite{ref4}).

In the case of a selfinteracting polymer chain the Hamiltonian has two
additional terms corresponding to the interactions between different
segments
\begin{eqnarray}
\beta{\cal H}_0(\br (n)) \longrightarrow & &
{\textstyle\frac{1}{2}}\beta\epsilon \int_{0}^{N} 
\ddot{\br}(n)^2~dn + {\textstyle\frac{1}{2}}\beta
\int_{0}^{N}\!\!\!\int_{0}^{N} dn dm~ V_{iso}\left( \br (n) - \br (m)\right)
+ \nonumber\\
& & + {\textstyle\frac{1}{2}}\beta \int_{0}^{N}\!\!\!\int_{0}^{N} dn dm \left(
\dot{\br}(n) \times \dot{\br}(m) \right)^2~ V_{an-iso}\left( \br (n) - \br
(m)\right), \nonumber\\
~
\end{eqnarray}
where the interaction has been split into an isotropic part and an
anisotropic part, of a general nematic form \cite{ref6}. 

After integrating out the variable $z(n)$ and taking into account the
fact that the chain is embedded into the surface $\zeta(\brho)$ one
obtains for the interaction part
\begin{equation}
\kern-50pt{\textstyle\frac{1}{2}}\beta \int_{0}^{N}\!\!\!\int_{0}^{N}
dn dm~ V_{iso}\left( \brho(n) - \brho(m) \right) 
~+~ {\textstyle\frac{1}{2}}\beta \int_{0}^{N}\!\!\!\int_{0}^{N} dn dm
~{\cal F}(\brho(n), \brho(m)) ~V_{an-iso}\left( \brho(n) - \brho(m)\right) 
\end{equation}
where I used the shorthand $ V\left( \brho(n) - \brho(m)\right) =
V \left( \brho(n) - \brho(m); \zeta(\brho(n)) - \zeta(\brho(m))
\right) $ and defined 
\begin{eqnarray}
\kern-70pt{\cal F}(\brho(n), \brho(m)) &&= \left( \dot{\brho}(n)^2 + \dot{\zeta}^2(\brho(n))\right)\left(
\dot{\brho}(m)^2 + \dot{\zeta}^2(\brho(m))\right) -
\left(\dot{\brho}(n)\dot{\brho}(m) +
\dot{\zeta}(\brho(n))\dot{\zeta}(\brho(m))\right)^2 = \nonumber\\
&& = \dot{\brho}^2_i(n)\dot{\brho}^2_k(m) -
\dot{\brho}_i(n)\dot{\brho}_i(m)\dot{\brho}_k(n)\dot{\brho}_k(m) +
\nonumber\\
&& + \left(
\dot{\brho}^2_i(m)\dot{\brho}_k(n)\dot{\brho}_l(n) +
\dot{\brho}^2_i(n)\dot{\brho}_k(m)\dot{\brho}_l(m) - 2
\dot{\brho}_i(n)\dot{\brho}_i(m)\dot{\brho}_k(n)\dot{\brho}_l(m)\right)
\frac{\partial\zeta}{\partial\brho_k}\frac{\partial\zeta}{\partial\brho_l}
+ \dots \nonumber\\
~
\end{eqnarray}
In the definition of ${\cal F}(\brho(n), \brho(m)$ I limited myself
only to terms up to and including the second order in the derivatives
of $\zeta(\brho)$, in order to be consistent with the harmonic
approximation assumed later for the bare surface Hamiltonian.

The final integration in the partition function has to be over the
surface degrees of freedom $\zeta(\brho)$. Assuming that in the
absence of the embedded polymer the surface energy is given by
$V(\zeta(\brho))$, the partition function can be cast into the form
\begin{equation}
\Xi (N) = \int\!\! \dots \!\!\int {\cal
D}\zeta(\brho)~\Xi(\zeta(\brho))\times exp(-\beta V(\zeta(\brho))). 
\label{equ6}
\end{equation}
In what follows I start from the assumption that $V(\zeta(\brho))$ is
a quadratic functional of the form
\begin{eqnarray}
V(\zeta(\brho)) &=& {\textstyle\frac{1}{2}}\gamma \int
\left(\nabla_{\bot}\zeta(\brho)\right)^2d^2\brho~ +~
{\textstyle\frac{1}{2}}K_c \int
\left(\nabla^2_{\bot}\zeta(\brho)\right)^2d^2\brho~ = \nonumber\\
&=& {\textstyle\frac{1}{2}}\int \gamma_{ik} \frac{\partial
\zeta(\brho)}{\partial \brho_{i}}\frac{\partial
\zeta(\brho)}{\partial \brho_{k}} d^2\brho +
{\textstyle\frac{1}{2}}\int K_{iklm} \frac{\partial^2 
\zeta(\brho)}{\partial \brho_{i}\partial \brho_{k}}\frac{\partial^2
\zeta(\brho)}{\partial \brho_{l}\partial \brho_{m}} d^2\brho,
\label{equ7}
\end{eqnarray}
where $\gamma$ is the bare surface tension, and $K_c$ is the bare
curvature elastic modulus. In order to facilitate later computations
the surface energy and the curvature modulus tensors are introduced,
which for a bare isotropic membrane assume the straightforward form
$\gamma_{ik} = \gamma \delta_{ik}$ and $K_{iklm} = K_c
\delta_{ik}\delta_{lm} $. By gathering all the terms in
$\ln{\Xi(\zeta(\brho))}$ that contribute to the quadratic order in
$\zeta(\brho)$ one will be in position to evaluate the rescaled values
of the elastic modulus and the surface tension, which is the aim of
this computation.

Also conveniently for later developements we introduce $ {\cal
S}_0({\bQ})$ through Fourier decomposed $\zeta(\brho)$
\begin{equation}
V(\zeta({\bQ})) = \sum_{\bQ} {\cal S}_0({\bQ})\vert \zeta({\bQ})\vert^2 ,
\label{equ7a}
\end{equation}
with
\begin{equation}
{\cal S}_0({\bQ}) = {\textstyle\frac{1}{2}} \gamma Q^2 +
{\textstyle\frac{1}{2}} K_c Q^4.
\end{equation}
${\cal S}_0({\bQ})$ is of course connected with the structure function
of the bare membrane.

\subsection{Collective Variables}

One proceeds by defining collective variables in terms of which the
partition function Eq.\ref{equ4} assumes the form that allows an
explicit introduction of mean - fields. The three collective variables
introduced to this effect are: density, orientational tensor and
the tensor of the squared curvature 
\begin{eqnarray}
\rho(\brho) &=& \int_0^N dn ~\delta(\brho - \brho(n)) \nonumber \\
\sigma_{ik}(\brho) &=& \int_0^N dn
~\dot{\brho}_i(n)\dot{\brho}_k(n)~\delta(\brho - \brho(n)) \nonumber
\\
c_{ik}(\brho) &=& \int_0^N dn
~\ddot{\brho}_i(n)\ddot{\brho}_k(n)~\delta(\brho - \brho(n)).
\label{equ8}
\end{eqnarray}
It should be noted here that since all the above collective variables
are defined on a 2D surface, they do not share the same properties
with their 3D counterparts. For instance the orientational tensor does
not satisfy the equality $Tr ~\sigma_{ik}(\brho) = \rho(\brho)$
\cite{ref6}. Later on an analogous relation valid in 2D will be
derived in detail. Furthermore the squared curvature tensor is not
connected with the torsion of the polymer chain but can be simplified by
means of the Frenet equation for a planar curve into
\begin{equation}
c_{ik}(\brho) = \int_0^N dn~
\kappa^2(n)~\bn_i(n)\bn_k(n)~\delta(\brho - \brho(n)) =  \mathopen<
\kappa^2(n) \bn_i(n)\bn _k(n)\mathclose>_{chain},
\end{equation}
where $\kappa(n)$ is the (planar) curvature and $\bn_i(n)$ is the
normal vector of the chain.

With collective variables the partition function Eq.\ref{equ4} can be
cast into the form
\begin{eqnarray}
\Xi (\zeta(\brho)) = & & \int\!\dots\!\int {\cal D}\brho(n) {\cal
D}\rho(\brho) {\cal 
D}\sigma_{ik}(\brho) {\cal D}c_{ik}(\brho)~\times \nonumber\\
& & \delta\left(\rho(\brho) - \int_0^N dn ~\delta(\brho - \brho(n))
\right) \times \nonumber\\
& & \delta\left(\sigma_{ik}(\brho) - \int_0^N dn
~\dot{\brho}_i(n)\dot{\brho}_k(n)~\delta(\brho - \brho(n)) \right)
\times \nonumber\\
& & \delta\left( c_{ik}(\brho) - \int_0^N dn
~\ddot{\brho}_i(n)\ddot{\brho}_k(n)~\delta(\brho - \brho(n))
\right)~\times \nonumber\\
& & \times~\exp{\left(-\beta H(\brho(n),\rho(\brho),\sigma_{ik}(\brho),c_{ik}(\brho))\right)}.  
\label{equ9}
\end{eqnarray}

The Hamiltonian $H(\brho(n), \rho(\brho), \sigma_{ik}(\brho),
c_{ik}(\brho))$ is obtained by substituting the collective variables
into ${\cal H}_0(\brho)$.  Assuming furthermore (for the sake of
convenience) that both the isotropic as well as the anisotropic parts
of the interaction potential have a form of a contact interaction
$V_{iso}(\brho,\brho') = u_{i}
\delta^2(\brho -\brho')$ and $V_{an-iso}(\brho,\brho') = u_{a}
\delta^2(\brho-\brho')$ one remains with the following form of the
interaction Hamiltonian
\begin{eqnarray}
\kern-70pt& &\beta H(\brho(n),\rho(\brho),\sigma_{ik}(\brho),c_{ik}(\brho)) =
{\textstyle\frac{1}{2}}\beta\epsilon \int_{0}^{N}\!\!dn~ \ddot{\brho}(n)^2
+ \lambda \int_{0}^{N}\!\!dn ~\dot{\brho}(n)^2~ + \nonumber\\
&+& {\textstyle\frac{1}{2}}\beta\epsilon \int\!\!d^2\brho ~\frac{\partial
\zeta(\brho)}{\partial \brho_i}\frac{\partial
\zeta(\brho)}{\partial \brho_k} ~c_{ik}(\brho) + \lambda
\int\!\!d^2\brho ~\frac{\partial \zeta(\brho)}{\partial \brho_i}\frac{\partial
\zeta(\brho)}{\partial \brho_k}~ \sigma_{ik}(\brho)~ + \nonumber\\
&+& ~{\textstyle\frac{1}{2}}\beta\epsilon \int\!\!d^2\brho~ \rho^{-1}(\brho)~
\frac{\partial^2 \zeta(\brho)}{\partial \brho_i \partial
\brho_k}\frac{\partial^2 \zeta(\brho)}{\partial \brho_l \partial
\brho_m} ~\sigma_{ik}(\brho) \sigma_{lm}(\brho) ~+ \nonumber\\
&+& \beta u_a \int d^2\brho \left[ \sigma_{ii}(\brho)
\sigma_{kl}(\brho) - \sigma_{ik}(\brho)\sigma_{il}(\brho) \right]
\frac{\partial\zeta}{\partial\brho_k}\frac{\partial\zeta}{\partial\brho_l}
~+ \nonumber\\
&+& {\textstyle\frac{1}{2}}\beta u_i \int d^2\brho ~\rho^2(\brho ) +
{\textstyle\frac{1}{2}}\beta u_a \int d^2\brho \left[
\sigma^2_{ii}(\brho ) - Tr \left(\sigma_{ik}(\brho )\right)^2 \right].  
\label{equ10}
\end{eqnarray}
Here only terms up to and including the second order in $\zeta(\brho)$
and its derivatives have been retained. A higher order term can be
derived from the $\ddot{\brho}_i(n)\dot{\brho}_k(n)\dot{\brho}_l(n)$
term in Eq.\ref{equ5} after the $\brho(n)$ integration and has been
consequently omitted.

\subsection{The Partition Function}

The next step is to introduce functional Fourier transform
representations for delta functions appearing in Eq.\ref{equ9}. For
each delta function we thus obtain an auxiliary field the integration
over which insures that the definitions in Eq.\ref{equ8} are
satisfied. Calling the auxiliary fields $\phi (\brho)$ for
$\rho(\brho)$, $\psi_{ik}(\brho)$ for $\sigma_{ik}(\brho)$ and
$t_{ik}(\brho)$ for $c_{ik}(\brho)$ we obtain the following relation
\begin{eqnarray}
\kern-100pt\Xi (\zeta(\brho)) &=& \int\!\dots\!\int {\cal D}\rho(\brho) {\cal
D}\sigma_{ik}(\brho) {\cal D}c_{ik}(\brho)~\times~{\cal D}\phi(\brho)
{\cal D}\psi_{ik}(\brho) {\cal D}t_{ik}(\brho) ~\times \nonumber\\
& & \Xi\left(\phi (\brho), \psi_{ik}(\brho),
t_{ik}(\brho)\right) ~\times~\exp{\left( -\beta{\cal
H}\left(\zeta(\brho); \phi (\brho), \psi_{ik}(\brho), t_{ik}(\brho);
\rho(\brho), \sigma_{ik}(\brho), c_{ik}(\brho)~ \right)\right)}, \nonumber\\
~ 
\label{equ11}
\end{eqnarray}
where 
\begin{equation}
\Xi \left(\phi (\brho), \psi_{ik}(\brho),
t_{ik}(\brho)\right) = \int\!\!\dots\!\!\int {\cal D}\brho(n)
~\exp{\left(-\beta{\cal H}_0\left( \brho(n); \phi (\brho), \psi_{ik}(\brho),
t_{ik}(\brho) \right)\right)}.
\end{equation}
The definitions employed in the above two formulas were
\begin{eqnarray}
\beta{\cal H}_0\left( \brho(n); \phi (\brho), \psi_{ik}(\brho),
t_{ik}(\brho) \right) = {\textstyle\frac{1}{2}}\beta\epsilon
\int_{0}^{N}\!\!dn~ \ddot{\brho}(n)^2 + \lambda \int_{0}^{N}\!\!dn
~\dot{\brho}(n)^2~ - \lambda \int_{0}^{N}\!\!dn + \nonumber\\ 
+~ \imathr \int_{0}^{N}\!\!dn~\phi(\brho(n)) + \imathr
\int_{0}^{N}\!\!dn~\psi_{ik}(\brho(n))
\dot{\brho}_i(n)\dot{\brho}_k(n) + \imathr
\int_{0}^{N}\!\!dn~t_{ik}(\brho(n)) \ddot{\brho}_i(n)\ddot{\brho}_k(n)
\nonumber\\
~
\label{equ12}
\end{eqnarray}
while 
\begin{eqnarray}
\kern+50pt& &\beta{\cal H}\left(\zeta(\brho); \phi (\brho), \psi_{ik}(\brho),
t_{ik}(\brho); \rho(\brho), \sigma_{ik}(\brho), c_{ik}(\brho)~ \right)
= \nonumber\\
&=& {\textstyle\frac{1}{2}}\beta\epsilon \int\!\!d^2\brho ~c_{ik}(\brho)~\frac{\partial
\zeta(\brho)}{\partial \brho_i}\frac{\partial
\zeta(\brho)}{\partial \brho_k}  + \lambda
\int\!\!d^2\brho ~\sigma_{ik}(\brho)~\frac{\partial \zeta(\brho)}{\partial \brho_i}\frac{\partial
\zeta(\brho)}{\partial \brho_k}~ + \nonumber\\
&+& ~{\textstyle\frac{1}{2}}\beta\epsilon \int\!\!d^2\brho~ \rho^{-1}(\brho)~
\sigma_{ik}(\brho) \sigma_{lm}(\brho)~\frac{\partial^2 \zeta(\brho)}{\partial \brho_i \partial
\brho_k}\frac{\partial^2 \zeta(\brho)}{\partial \brho_l \partial
\brho_m}~ + \nonumber\\
&+& \beta u_a \int d^2\brho \left[ \sigma_{ii}(\brho)
\sigma_{kl}(\brho) - \sigma_{ik}(\brho)\sigma_{il}(\brho) \right]
\frac{\partial\zeta}{\partial\brho_k}\frac{\partial\zeta}{\partial\brho_l}
~+ \nonumber\\
&+& {\textstyle\frac{1}{2}}\beta u_i \int d^2\brho ~\rho^2(\brho ) +
{\textstyle\frac{1}{2}}\beta u_a \int d^2\brho \left[
\sigma^2_{ii}(\brho ) - Tr \left(\sigma_{ik}(\brho )\right)^2
\right] ~- \nonumber\\
&-& ~ \imathr \int\!\!d^2\brho ~\rho(\brho)\phi(\brho) - \imathr
\int\!\!d^2\brho ~\sigma_{ik}(\brho) \psi_{ik}(\brho) - \imathr
\int\!\!d^2\brho ~ c_{ik}(\brho)t_{ik}(\brho). \nonumber\\
~
\label{equ13}
\end{eqnarray}

The partition function of the membrane with an embedded flexible
polymer chain can now be obtained in several straightforward
steps. Let us first of all introduce the renormalized potential
$\tilde{V}(\zeta(\brho))$ which is of the same form as Eq.\ref{equ7}
but with renormalized values of $\gamma_{ik}$ and $K_{iklm}$
\begin{eqnarray}
\gamma_{ik}(\brho) &\longrightarrow& \gamma \delta_{ik} + \epsilon
c_{ik}(\brho) + 2\frac{\lambda}{\beta} \sigma_{ik}(\brho) + 2 u_a \left[\sigma_{ll}(\brho)
\sigma_{ik}(\brho) - \sigma_{li}(\brho)\sigma_{lk}(\brho)\right] \nonumber\\
K_{iklm}(\brho) &\longrightarrow& K_c \delta_{ik}\delta_{lm} +
\frac{\epsilon}{\rho} (\brho) \sigma_{ik}(\brho) \sigma_{lm}(\brho). \nonumber\\
~
\label{equ14}
\end{eqnarray}
The renormalized values of $\gamma_{ik}$ and $K_{iklm}$ thus depend on
the set $\rho(\brho), \sigma_{ik}(\brho)$ and $c_{ik}(\brho)$.  The
corresponding Fourier representation would ensue as
\begin{equation}
\tilde{V}(\zeta(\bQ)) = {\textstyle\frac{1}{2}} \sum_{\bQ}\left(
\gamma_{ik}(\bQ) Q_{i}Q_{k} + K_{iklm}(\bQ)Q_{i}Q_{k}Q_{l}Q_{m} \right)
\vert \zeta({\bQ})\vert^2. 
\end{equation}

Now we construct a generating functional (functional Fourier
transform) of $ \exp{-\beta\tilde{V} (\zeta(\brho),
\rho(\brho), \sigma_{ik}(\brho), c_{ik}(\brho))}$ at some quenched
value of the membrane shape $\zeta(\brho) $. By definition one has
\begin{eqnarray}
\kern-70pt\exp{\left(-\beta\tilde{V}(\zeta(\brho),\phi(\brho),
\psi_{ik}(\brho), t_{ik}(\brho))\right)} &=& {\cal
Z}_{\zeta(\brho)}\left( \phi(\brho),
\psi_{ik}(\brho), t_{ik}(\brho) \right) =
\int\!\!\dots\!\!\int {\cal D}\rho(\brho) {\cal D}\sigma_{ik}(\brho)
{\cal D}c_{ik}(\brho) \times \nonumber\\ & & 
\kern-200pt~\times\exp{\left(-\beta\tilde{V}(\zeta(\brho),
\rho(\brho), \sigma_{ik}(\brho), c_{ik}(\brho)) - \imathr 
\int\!\!d^2\brho ~\rho(\brho)\phi(\brho) - \imathr 
\int\!\!d^2\brho ~\sigma_{ik}(\brho) \psi_{ik}(\brho) - \imathr
\int\!\!d^2\brho ~ c_{ik}(\brho)t_{ik}(\brho)\right)}, \nonumber\\
~
\label{equ15}
\end{eqnarray}
where the l.h.s. should be read symbolically as a shorthand for the
functional Fourier transform.

Finally the partition function can be expressed as an average of the
product of the generating functional with variables $\phi(\brho),
\psi_{ik}(\brho), t_{ik}(\brho)$ and the partition function of a
single polymer chain in the orienting field of external sources 
$\phi (\brho), \psi_{ik}(\brho)~$ and $~t_{ik}(\brho)$,
averaged over all the conformations of the membrane, {\sl i.e.}
\begin{equation}
\Xi (N) = \int {\cal D}\zeta(\brho) {\cal Z}(\zeta(\brho))
\end{equation}
with
\begin{equation}
\kern-60pt{\cal Z}(\zeta(\brho)) = 
\int\!\!\dots\!\!\int {\cal D}\phi(\brho) {\cal D}\psi_{ik}(\brho)
{\cal D}t_{ik}(\brho)~ \exp{\left(-\beta\tilde{V}(\zeta(\brho),\phi(\brho),
\psi_{ik}(\brho), t_{ik}(\brho))\right)} ~
\Xi \left(\phi (\brho), \psi_{ik}(\brho), t_{ik}(\brho)\right). 
\label{equ16}
\end{equation}
Obviously the general form of the partition function is untractable
and additional approximations have to be considered to get a
closed-form solution.

\subsection{Mean-field Approximation}

It was recently shown \cite{ref1} that a similar partition function can
be evaluated explicitely by introducing a mean - field description of
the polymer collective coordinates $ \rho(\brho),
\sigma_{ik}(\brho), c_{ik}(\brho)$ as well as their auxiliary fields
$ \phi(\brho), \psi_{ik}(\brho), t_{ik}(\brho) $  by supressing their
$\brho $  dependence and treating them as constants. Thus by
introducing  $ \rho(\brho) \rightarrow \bar{\rho}, \sigma_{ik}(\brho)
\rightarrow \bar{\sigma}_{ik}, c_{ik}(\brho) \rightarrow \bar{c}_{ik}
$ as well as $ \dot{\imath}\phi(\brho) \rightarrow \bar{\phi},
~\dot{\imath}\psi_{ik}(\brho) \rightarrow \bar{\psi}_{ik},
~\dot{\imath}t_{ik}(\brho) \rightarrow \bar{t}_{ik} $ one can first of
all explicitely evaluate the partition function of a single chain with
external sources $ \Xi \left(\bar{\phi}, \bar{\psi}_{ik},
\bar{t}_{ik}\right)$. This calculation will not be repeated as it has
been already detailed in \cite{ref1}. The final result valid in the
limit $N \gg 1$ can be obtained as
\begin{equation}
- kT \ln{\Xi \left(\bar{\phi}, \bar{\psi}_{ik}, \bar{t}_{ik}\right)} =
{\textstyle\frac{N}{2}}\sum_{\alpha} \sqrt{\frac{\left( \lambda +
\bar{\psi}_{\alpha}\right) }{\left({\textstyle \frac{\beta\epsilon}{2}} +
\bar{t}_{\alpha}\right)}}~+~{\textstyle \frac{N}{2\pi}}q_{\tiny
max}\sum_{\alpha} \ln{\left(1 + \frac{2\bar{t}_{\alpha}}{\beta
\epsilon}\right)}~-~N\lambda~+~N\bar{\phi}, 
\label{equ17}
\end{equation}  
where index $\alpha$ refers to the eigenvalues so that
$\bar{\psi}_{\alpha}, \bar{t}_{\alpha}$ are the indexed eigenvalues of
$\bar{\psi}_{ik}$ and $\bar{t}_{ik}$. The next step now is to evaluate
${\cal Z}_{\zeta(\brho)}\left( \bar{\phi}, \bar{\psi}_{ik},
\bar{t}_{ik}\right)$ on the mean - field level. In complete analogy
with the calculations in \cite{ref1} I obtain
\begin{eqnarray}
\kern-40pt - kT \ln{{\cal Z}_{\zeta(\brho)}\left( \bar{\phi}, \bar{\psi}_{ik},
\bar{t}_{ik}\right)} &=& {\textstyle\frac{1}{2}}\beta \epsilon
~\bar{c}_{\alpha} 
\int\!\!d^2\brho ~\left(\frac{\partial \zeta(\brho)}{\partial
\brho_{\alpha}}\right)^2 + \lambda ~\bar{\sigma}_{\alpha} 
\int\!\!d^2\brho ~\left(\frac{\partial \zeta(\brho)}{\partial
\brho_{\alpha}}\right)^2 ~ + \nonumber\\
&+& ~{\textstyle\frac{1}{2}}\beta \epsilon ~\bar{\rho}^{\tiny -1}
~\bar{\sigma}_{\alpha}\bar{\sigma}_{\beta} \int\!\!d^2\brho~
\frac{\partial^2 \zeta(\brho)}{\partial
\brho_{\alpha}^2} \frac{\partial^2 \zeta(\brho)}{\partial
\brho_{\beta}^2}~+ \nonumber\\ 
&+& \beta u_a \left({\rm Tr}\bar{\sigma}_{\alpha}\right) \bar{\sigma}_{\beta}\int
d^2\brho \left(\frac{\partial\zeta(\brho)}{\partial\brho_{\beta}}\right)^2 -
\beta u_a \bar{\sigma}_{\alpha}\bar{\sigma}_{\beta} \int d^2\brho
\frac{\partial\zeta(\brho)}{\partial\brho_{\alpha}}\frac{\partial\zeta(\brho)}{\partial\brho_{\beta}}~+
\nonumber\\
&+& {\textstyle\frac{1}{2}}\beta u_i S ~\bar{\rho}^2
~+~{\textstyle\frac{1}{2}}\beta u_a S \left[ \left({
Tr} \bar{\sigma}_{\alpha}\right)^2 - {Tr}\left(\bar{\sigma}_{\alpha}\right)^2 \right] ~- \nonumber\\
&-& ~ S~\bar{\rho}\bar{\phi} - S~\bar{\sigma}_{\alpha}\bar{\psi}_{\alpha} - S~\bar{c}_{\alpha}\bar{t}_{\alpha}. 
\label{equ18}
\end{eqnarray}
What remains now is the final integration over the membrane modes
$\zeta(\brho)$ or equivalently over their Fourier components
$\zeta(\bQ)$. It is performed in such a way that all the terms
describing the bare membrane and thus depending only on ${\cal
S}_0(\bQ)$ are discarded as they do not enter the mean - field
evaluation of collective variables or their auxiliary fields. The
final expression obtained in this way for $\Xi (N)$ thus assumes the form
\begin{eqnarray}
\kern-90pt- kT \ln{\Xi (N)} &\cong& {\textstyle \frac{N}{2}}\sum_{\alpha}
\sqrt{\frac{\left( \lambda + \bar{\psi}_{\alpha}\right)
}{\left({\textstyle \frac{\beta \epsilon}{2}} + 
\bar{t}_{\alpha}\right)}}~+~{\textstyle \frac{N}{2\pi}}q_{\tiny
max}\sum_{\alpha} \ln{\left(1 + \frac{2\bar{t}_{\alpha}}{\beta
\epsilon}\right)} - N\lambda + N\bar{\phi} - \nonumber\\
&-& S~\bar{\rho}\bar{\phi} -
S~\sum_{\alpha} \bar{\sigma}_{\alpha}\bar{\psi}_{\alpha} -
S~\sum_{\alpha} \bar{c}_{\alpha}\bar{t}_{\alpha} +
{\textstyle\frac{1}{2}}\beta u_i S ~\bar{\rho}^2 + 
{\textstyle\frac{1}{2}}\beta u_a S \left[ \left({ 
Tr}\bar{\sigma}_{\alpha}\right)^2 - {
Tr}\left(\bar{\sigma}_{\alpha}\right)^2 \right] ~+~ \nonumber\\
&+& {\textstyle \frac{1}{2}} \sum_{\bQ} \ln{\left( 1 + \beta\epsilon
\sum_{\alpha} \bar{c}_{\alpha} \frac{Q_{\alpha}^2}{\beta 
{\cal S}_0(\bQ)} + 2\lambda \sum_{\alpha} \bar{\sigma}_{\alpha} 
\frac{Q_{\alpha}^2}{\beta {\cal S}_0(\bQ)} + \frac{\beta
\epsilon}{\bar{\rho}}\sum_{\alpha,\beta}
\bar{\sigma}_{\alpha}\bar{\sigma}_{\beta} 
\frac{Q_{\alpha}^2 Q_{\beta}^2}{\beta {\cal S}_0(\bQ)} 
\right)}. \nonumber\\
~
\label{equ19}
\end{eqnarray}
In the above equation I have limited myself to the lowest orders in
$\bar{c}_{\alpha}$ and $\bar{\sigma}_{\alpha}$, meaning to the first
order in $Q^2$ terms and to the second order in
$\bar{\sigma}_{\alpha}$ for $Q^4$ terms, all the higher orders have
been discarded. This essentially limits the validity of conclusion
derived from Eq.\ref{equ19} to the regime of low surface coverage of
the membrane, {\sl i.e.} to small $\bar{\rho}$.

Introducing now 
\begin{equation}
\beta{\cal S}({\bQ}) = \beta{\cal S}_0(\bQ) + \beta\epsilon
\sum_{\alpha} \bar{c}_{\alpha} Q_{\alpha}^2  + 2\lambda \sum_{\alpha}
\bar{\sigma}_{\alpha} Q_{\alpha}^2 + \frac{\beta
\epsilon}{\bar{\rho}}\sum_{\alpha,\beta} \bar{\sigma}_{\alpha}\bar{\sigma}_{\beta} 
Q_{\alpha}^2 Q_{\beta}^2 
\end{equation}
I obtain the following set of mean-field equations (the leftmost
quantity stands for the minimizing variable)
\begin{eqnarray}
\bar{c}_{\alpha} &:& \bar{t}_{\alpha} = \frac{\beta\epsilon}{2}
\sum_{\bQ} \frac{Q_{\alpha}^2}{\beta {\cal S}(\bQ)} \nonumber\\
\bar{t}_{\alpha} &:& \bar{c}_{\alpha} = - {\textstyle \frac{N/S}{4}} 
\sqrt{\frac{\left( \lambda + \bar{\psi}_{\alpha}\right)
}{\left({\textstyle \frac{\beta \epsilon}{2}} +
\bar{t}_{\alpha}\right)}}\left( {\textstyle \frac{\beta \epsilon}{2}} +
\bar{t}_{\alpha} \right)^{-1} + {\textstyle
\frac{N/S}{\pi}}\frac{q_{max}}{\beta\epsilon} \left( 1 +
\frac{2\bar{t}_{\alpha}}{\beta\epsilon}\right)^{-1} \nonumber\\
\bar{\sigma}_{\alpha} &:& \bar{\psi}_{\alpha} = \beta u_a \left[ \left({
Tr}\bar{\sigma}_{\alpha}\right) - \bar{\sigma}_{\alpha}\right] +
\lambda \bar{\sigma}_{\alpha} \sum_{\bQ} \frac{Q_{\alpha}^2}{\beta {\cal
S}(\bQ)} +  \frac{\beta\epsilon}{\bar{\rho}}\sum_{\beta}
\bar{\sigma}_{\beta} \sum_{\bQ}\frac{Q_{\alpha}^2 Q_{\beta}^2}{\beta
{\cal S}(\bQ)} \nonumber\\
\bar{\psi}_{\alpha} &:& \bar{\sigma}_{\alpha} = {\textstyle
\frac{N/S}{4}} \frac{1}{\sqrt{\left( \lambda +
\bar{\psi}_{\alpha}\right)\left({\textstyle \frac{\beta \epsilon}{2}}
+ \bar{t}_{\alpha}\right)}} \nonumber\\
\bar{\rho} &:& \bar{\phi} = \beta u_i \bar{\rho} - \frac{\beta
\epsilon}{2\bar{\rho}^2}\sum_{\alpha,\beta} \sum_{\bQ}
\bar{\sigma}_{\alpha}\bar{\sigma}_{\beta} 
\frac{Q_{\alpha}^2 Q_{\beta}^2}{\beta {\cal S}(\bQ)} \nonumber\\
\bar{\phi} &:& \bar{\rho} = N/S \nonumber\\
\lambda &:& {\textstyle \frac{N/S}{4}} \sum_{\alpha} \frac{1}{\sqrt{\left( \lambda +
\bar{\psi}_{\alpha}\right)\left({\textstyle \frac{\beta \epsilon}{2}}
+ \bar{t}_{\alpha}\right)}} = (N/S) -
\sum_{\alpha}\bar{\sigma}_{\alpha} 
\sum_{\bQ} \frac{Q_{\alpha}^2}{\beta {\cal S}(\bQ)}. 
\end{eqnarray}
The appearance of wave-vector $q_{max}$ is a consequence of the limit of
the continuous description of a polymer chain and should be inversely
proportional to the Kuhn's length.

At this point a perturbation expansion in terms of the surface
coverage of the polymers $\bar{\rho}$ is introduced and only the
lowest order is taken into account, essentially making the
approximation ${\cal S}(\bQ) \longrightarrow {\cal S}_0(Q)$. Since
${\cal S}_0(Q)$ is by assumption isotropic one has
\begin{eqnarray}
\sum_{\bQ} \frac{Q_{\alpha}^2}{\beta {\cal S}(\bQ)} &\cong& \sum_{\bQ}
\frac{Q_{\alpha}^2}{\beta {\cal S}_0(Q)} = \lim_{\brho \rightarrow
\brho'} \left< \frac{\partial \zeta(\brho)}{\partial
\brho_\alpha}\frac{\partial \zeta(\brho')}{\partial \brho'_\alpha}
\right>_0 = F \nonumber\\
\sum_{\bQ} \frac{Q_{\alpha}^2Q_{\beta}^2}{\beta {\cal S}(\bQ)} &\cong&
\sum_{\bQ} \frac{Q_{\alpha}^2Q_{\beta}^2}{\beta {\cal S}_0(Q)} = 
\lim_{\brho \rightarrow \brho'} 
\left< \frac{\partial^2 \zeta(\brho)}{\partial\brho^2_\alpha}
\frac{\partial^2 \zeta(\brho')}{\partial \brho'^2_\beta} 
\right>_0= F_{||} \delta_{\alpha\beta} + F_{\perp}\left( 1 -
\delta_{\alpha\beta}\right). \nonumber\\
~
\end{eqnarray}
Index $0$ signifies that the averages are taken with respect to a bare
membrane. Obviously $F$ is the strength of the surface normal
fluctuations and $ F_{||}$ and $F_{\perp}$ are the strengths of the
surface curvature fluctuations.

At this point the calculation performed here substantially deviates
from \cite{ref1}. Since in that case the polymer was ideal, {\sl i.e.}
non-interacting, the orientational ordering was due to higher order
expansion terms in ${\cal S}(\bQ)$, omitted here, as they would only
rescale the already existing transition point but not its qualitative
characteristics.

The mean-field equations can now be solved by means of the {\sl
ansatz}
\begin{equation}
\bar{\sigma}_{\alpha} = \frac{\bar{\rho}}{2(1 + F)}( 1 \pm S),
\end{equation}
where $S$ is the orientational order parameter $ S = \frac{1}{2}\left(
3 \mathopen<\cos^2{\Theta}\mathclose>- 1 \right)$ \cite{prost} with
$\Theta$ being the angle between the local direction and the nematic
axis. It thus follows that in this case ${Tr}~\bar{\sigma}_{ik} =
{\bar{\rho}}/{(1 + F)}$ and as announced, thus differs from the 3D
equivalent ${Tr}~\bar{\sigma}_{ik} = \bar{\rho}$. The difference is
due to the projection of the continuity condition Eq. \ref{equ5} onto
a fluctuating surface \cite{ref1}.

Putting this {\sl ansatz} into the minimization conditions we are left
with the following results
\begin{eqnarray}
( 1 - S^2)^2 &=& \frac{1}{2(\beta\epsilon\bar{\rho})\left( \beta u_a -
\frac{\beta\epsilon}{\bar{\rho}}(F_{||} - F_{\perp})\right)}
\nonumber\\
\lambda &=& \frac{1}{2(\beta\epsilon)}\frac{2(1 + S^2)}{(1 - S^2)^2} -
\frac{\bar{\rho}}{(1 + F)^2}\left( \beta u_a +
\frac{\beta\epsilon}{\bar{\rho}}(F_{||} + F_{\perp})\right)
\nonumber\\
\bar{c}_{\alpha} &=& \frac{\bar{\rho}}{2(\beta\epsilon)^2(1 +
F)}\frac{(1 \mp S)}{(1 - S^2)} +
\frac{2\bar{\rho}}{\beta\epsilon}\frac{q_{max}}{2\pi(1 + F)}.
\end{eqnarray}
Clearly the solution of the above equations describes a second order
surface orientational transition characterized by $S=0$ and $S \neq
0$. The critical point of the transition is given as a solution of
\begin{equation}
2\beta\epsilon\bar{\rho} = \frac{1}{\left( \beta u_a -
\frac{\beta\epsilon}{\bar{\rho}}(F_{||} - F_{\perp})\right)}.
\end{equation}
The transition involves only the variables $\beta, \epsilon$ and
$\bar{\rho}$, and can thus be achieved by either a variation in
temperature, polymer stiffness or polymer surface coverage.

\subsection{Effective Elastic Constants}

The corresponding renormalization of the effective membrane elastic
constants can be obtained on this level of approximations from
\begin{eqnarray}
{\cal S}(\bQ) &\longrightarrow& {\cal S}_0(Q) + \beta\epsilon
\sum_{\alpha} \bar{c}_{\alpha} Q_{\alpha}^2  + 2\lambda \sum_{\alpha}
\bar{\sigma}_{\alpha} Q_{\alpha}^2 + \frac{\beta
\epsilon}{\bar{\rho}}\sum_{\alpha,\beta} \bar{\sigma}_{\alpha}\bar{\sigma}_{\beta} 
Q_{\alpha}^2 Q_{\beta}^2 = \nonumber\\ 
&=& {\textstyle\frac{1}{2}}\beta\gamma Q^2 +
{\textstyle\frac{1}{2}}\beta K_c Q^4 + {\textstyle\frac{1}{2}}\beta\epsilon
\sum_{\alpha} \bar{c}_{\alpha} Q_{\alpha}^2  + \lambda \sum_{\alpha}
\bar{\sigma}_{\alpha} Q_{\alpha}^2 + {\textstyle\frac{1}{2}}\frac{\beta
\epsilon}{\bar{\rho}}\sum_{\alpha,\beta} \bar{\sigma}_{\alpha}\bar{\sigma}_{\beta} 
Q_{\alpha}^2 Q_{\beta}^2. \nonumber\\
~
\end{eqnarray} 
The renormalized values of the elastic constants can be read off as
\begin{eqnarray}
\gamma &\longrightarrow& \gamma + 2\frac{\lambda}{\beta} \bar{\sigma}_{\alpha} + \epsilon \bar{c}_{\alpha} \nonumber\\
K_c &\longrightarrow& K_c +
\frac{\epsilon}{\bar{\rho}}\bar{\sigma}_{\alpha}\bar{\sigma}_{\beta}.
\end{eqnarray}
One observes first of all that the elastic modulus of the ``dressed''
membrane changes because of the elasticity of the embedded polymer,
while the surface tension changes because of two distinct
contributions: one stemming from the fact that polymer is embedded
into the surface and thus scales linearly with $\lambda$, the other
one stemming from the coupling between the curvature of the surface
and curvature of the embedded chain.

If I now write 
\begin{eqnarray}
\beta{\cal S}(Q) &=& {\textstyle\frac{1}{2}}\Sigma_{||}\left( Q_x^2 + Q_y^2
\right) + {\textstyle\frac{1}{2}}\Sigma_{\perp}S\left( Q_x^2 -
Q_y^2\right) + \nonumber\\
& & + {\textstyle\frac{1}{2}}\beta K_c\left( Q_x^2 +
Q_y^2\right)^2 +
{\textstyle\frac{1}{2}}\frac{\beta\epsilon}{\rho}\left(\sigma_{||}\left(
Q_x^2 + Q_y^2\right) + \sigma_{\perp}S\left( Q_x^2 - Q_y^2\right)\right)^2
\nonumber\\
~
\label{equcit}
\end{eqnarray}
thus defining the quantities $\sigma_{||}$ and $\sigma_{\perp}$ as
well as $\Sigma_{||}$ and $\Sigma_{\perp}$, I obtain
\begin{eqnarray}
\Sigma_{||} &=& \beta\gamma + 2\beta\epsilon~c_{||} + 2\left(2\lambda +
\beta u_a\frac{\bar{\rho}}{(1+F)}\right)\sigma_{||} \nonumber\\
\Sigma_{\perp} &=& 2 \beta\epsilon~c_{\perp} + 2\left(2\lambda +
\beta u_a\frac{\bar{\rho}}{(1+F)}\right)\sigma_{\perp},
\end{eqnarray}
while
\begin{eqnarray}
\sigma_{||}  &=& \sigma_{\perp} = \frac{\bar{\rho}}{2(1+F)} \nonumber\\
c_{||}  &=& \left(
\frac{2\bar{\rho}}{\beta\epsilon}\frac{q_{max}}{2\pi(1 + F)} -
c_{\perp}\right) \nonumber\\
c_{\perp} &=& \frac{\bar{\rho}}{2(\beta\epsilon)^2(1+F)(1-S^2)}.
\label{definitions}
\end{eqnarray}
It is clear from Eq. \ref{equcit} that depending on the orientational
order parameter in the nematic phase the elastic energy will be
destabilized if $\Sigma_{||} - \Sigma_{\perp}S < 0$. This
destabilization of the surface energy is due to the existence of an
easy axis, in the direction perpendicular to the nematic axis,
characterized by a much smaller surface energy than in the direction
parallel to the nematic axis. This effect follows directly from the
anysotropic packing of the polymer on the embedding surface.

\subsection{The Tubulization Transition}

Renormalized elastic constants are now taken as a point of departure
for the general analysis of the effective elastic energy. The recent
quite general formalism developed by Radzihovsky and Toner \cite{ref7}
is taken as a lead.

Instead of choosing the Monge parametrization for the surface ${\bf
r}(x,y) = (x,y, \zeta(x,y))$ one can write down the elastic energy
Eq. \ref{equcit} directly in terms of ${\bf r}(x_i)$ with $x_1 = x,
x_2 = y$. Also in order to insure stability, in view of the discussion
preceeding this section, appropriate fourth order terms \cite{maya}
have to be added to the expansion Eq. \ref{equcit}. In the spirit of
the Landau theory we assume that the fourth order displacement terms
do not change sign as a function of the nematic order parameter, and
thus assume them as constant.

The total elastic energy, meaning the sum of Eq. \ref{equcit} and the
fourth order terms, can now be written in terms of the gradients of
local displacements as \cite{maya}, 
\begin{eqnarray}
\kern-70pt{\cal F}({\bf r}(x_i)) &&= {\textstyle\frac{1}{2}} \int dx_{\perp} dx_{n}
\left[ K_{\perp\perp}\left( \partial^2_{\perp}{\bf r}(x_i) \right)^2 +
K_{nn}\left( \partial_{n}^2{\bf r}(x_i) \right)^2 \right. +
K_{n \perp} \left(  \partial_{\perp}^2{\bf r}(x_i)\right) \left(\partial^2_{n}
{\bf r}(x_i) \right) + \nonumber\\
&& +~ t_{\perp}\left( \partial_{\perp}{\bf r}(x_i) \right)^2 +
t_{n}\left( \partial_{n}{\bf r}(x_i) \right)^2 + \nonumber\\
&& +~ {\textstyle\frac{1}{2}} u_{\perp\perp} \left( \partial_{\perp}{\bf
r}(x_i) . \partial_{\perp}{\bf r}(x_i) \right)^2 + {\textstyle\frac{1}{2}}
u_{nn}  \left(\partial_{n}{\bf r}(x_i)
. \partial_{n}{\bf r}(x_i) \right)^2 + u_{n\perp}
\left(  \partial_{n}{\bf r}(x_i)  . \partial_{\perp}{\bf
r}(x_i)\right)^2 + \nonumber\\
&& +~  {\textstyle\frac{1}{2}} v_{\perp \perp } 
\left( \partial_{\perp}{\bf r}(x_i) . \partial_{\perp}{\bf r}(x_i) \right)^2 + 
v_{n\perp} \left(  \partial_{n}{\bf r}(x_i)\right)^2
\left( \partial_{\perp}{\bf r}(x_i)\right)^2 \left. \right],
\end{eqnarray}
where we have chosen the nematic direction to be $n$.  In order for
the system to be stable the fourth order constants have to be positive
$u, v > 0$. The second order constants can be obtained by comparison
with Eq. \ref{equcit}. The effective elastic moduli are obtained as
\begin{eqnarray}
K_{\perp\perp} &&= K_c + \frac{\beta\epsilon}{\rho}\left(
\sigma_{n}^2 -  \sigma_{n}\sigma_{\perp}S +  \sigma_{\perp}^2S^2 \right) \nonumber\\
K_{nn} && = K_c +   \frac{\beta\epsilon}{\rho}\left(
\sigma_{n}^2 +  \sigma_{n}\sigma_{\perp}S +  \sigma_{\perp}^2S^2 \right)\nonumber\\
K_{n \perp} && = K_c + \frac{\beta\epsilon}{\rho}\left(
\sigma_{n}^2 - \sigma_{\perp}^2S^2 \right),
\end{eqnarray}
and as can be easily deduced from the definitions Eq. \ref{definitions}
are all positive. The effective surface tension can be cast into the form
\begin{eqnarray}
t_{n} &&= \Sigma_{n} + \Sigma_{\perp}S \nonumber\\ 
t_{\perp} &&= \Sigma_{n} - \Sigma_{\perp}S,
\end{eqnarray}
where $t_{\perp}$ can be either positive or negative. The zero of
$t_{\perp}$ does not coincide with the isotropic - nematic transition
of the polymers, {\sl i.e. } with $S = 0$.

This free energy is now treated in analogy with the usual $\phi^4$
theories of the critical phenomena, identifying the tangent vectors
${\bf t}_{\alpha} = \partial_{\alpha}{\bf r}(x_i)$ as order parameters
\cite{maya}. In what follows we will delimit the analysis to the
mean-field approximation where we take the {\sl ansatz} ${\bf r}_{MF}
= \left( \zeta_{\perp} x, \zeta_{n} y
\right)$, where the symmetry breaking nematic axis has been oriented
in the $y$ (parallel) direction. The prefactors $\zeta{\perp},
\zeta_{n}$ are order parameters that measure the shrinkage of
the membrane due to undulations \cite{david}.  With this {\sl ansatz}
one obtains for the free energy
\begin{equation}
{\cal F} = {\textstyle\frac{1}{2}} L_{\perp}.L_{n} \left[
\right.  t_{n}\zeta_{n}^2 + t_{\perp} \zeta_{\perp}^2
+ {\textstyle\frac{1}{2}}\left( u_{\perp\perp} + v_{\perp\perp}
\right)\zeta_{\perp}^4 + {\textstyle\frac{1}{2}}u_{nn}
\zeta_{n}^4 + v_{n\perp} \zeta_{n}^2
\zeta_{\perp}^2 \left. \right].
\end{equation}
Following Radzihovsky and Toner \cite{ref7} by minimizing the above
free energy one obtains the following phase diagram topology for
$\left( u_{\perp\perp} + v_{\perp\perp} \right)u_{nn}
>  v_{n\perp}^2$. There exists a crumpled phase with $
\zeta_{n} = \zeta_{\perp} = 0$ for $t_{n},  t_{\perp}
> 0$. As $t_{\perp}$ changes sign a tubular phase sets in  with $
\zeta_{\perp} = \sqrt{\frac{\vert t_{\perp} \vert}{u_{\perp\perp}}}$
and $\zeta_{n} = 0$. This transition is second order. A
transition to the flat phase, which can be expected on general grounds
\cite{ref7}, never sets in as $t_{n}$ can not change sign.

That $ t_{\perp}$ is in deed bound to change sign somewhere within the
polymer surface nematic phase is evident from the fact that
$\lim_{S\longrightarrow 0} t_{\perp} = \Sigma_{n} \sim const. > 0$
while in the opposite limit $\lim_{S\longrightarrow 1} t_{\perp} =
\Sigma_{n} - \Sigma_{\perp} \sim - (1 - S^2)^{\textstyle -1} < 0$. The
exact position of the change of sign does not coincide with the
isotropic - nematic transition of the embedded polymer chain, {\sl
i.e.} with $S = 0$. Thus a sufficient ordering of the polymer chain
has to be present before the tubulization transition of the whole
embedding membrane can take place.

A second order orientational transition of the embedded polymer chain
thus drives an associated shape transition of the membrane,
corresponding to a tubulization of the membrane in the direction
perpendicular to the nematic axis, or parallel to the easy axis of the
surface energy. The two transitions ({\sl i.e.} the nematic to
isotropic transition of the polymer and the symmetry breaking shape
transition of the dressed membrane) do not, however, happen at the
same value of the temperature (surface coverage or polymer stiffness).

\section{Discussion}

Strong adsorption of polymers onto soft supporting surfaces has
different characteristics from ordinary polymer adsorption where the
adsorption phenomenon results from a competition between adsorption
energy and entropically driven tendency for less tightly bound
configurations. Recent experimental work \cite{ref2,ref3} shows that in
the strong adsorption case the polymer chains are basically confined
to lie on the adsorbing surface, without any dangling end or train
configurations. In this case, with polymer chains effectively embedded
(onto) into the supporting surface, the main source of entropy is
limited to either polymer conformational degrees of freedom along the
embedding surface or to thermally driven fluctuations of the
supporting surface.

Because of this tight coupling of the polymer with the supporting
surface, polymer statistics is strongly influenced by local surface
configurations. Furthermore, if the polymer is non-ideal, {\sl i.e.}
self-interacting, the supporting surface fluctuations renormalize the
interactions between polymer segments as has been amply demonstrated
in a slightly different context by Goulian {\sl et al.}
\cite{mark}. This effect depends on the nature of the polymer
self-interactions as well as on the magnitude of the supporting
surface fluctuations. We have investigated the properties and
phenomena connected with an orientational ordering transition of
surface-embedded polymers exhibiting an orientational (nematic)
interaction potential and have shown explicitely the nature of the
renormalization of the polymer self-interaction due to the supporting
surface fluctuations. The conclusion reached in this connection is
that for a sufficiently stiff membrane this effect tends to be small
since it depends on the equilibrium fluctuations of the bare
surface curvature.

It is however not only true that thermally driven undulations of the
supporting surface affect the polymer ordering transition, but this
transition in its turn, manifestly modifies the properties of the
compound (bare membrane + embedded polymers) membrane. The reverse
effect, of polymer modified membrane properties, appears to be much
more important both in qualitative and quantitative terms. Polymer
orientational ordering spawns an associated symmetry breaking
transition in the mean shape of the compound membrane that preferrs
membrane configurations with most of the surface area in the direction
of the easy (low energy) axis of the surface energy. This is achieved
through a tubulization transition where the long axis of the tubule
is perpendicular to the nematic axis of the polymer ordering.

Though it is understandable that nematic ordering of a surface
embedded polymer should have some consequences on the properties of
the ``dressed'' surface, the drastic change in the mean shape of the
compound membrane does come as a surprise.

One can foresee several effects that are beyond the presently
formulated mean-field solution to the tight adsorption model. First of
all the coupling between local curvatures and the nematic order
parameter of the polymer $S$ is missing. It is intuitively plausible
and in deed probable, that local curvature should affect local
ordering of the polymer segments tending to concentrate alligned
polymers in regions with large local curvature. This effect could also
introduce additional ordered phases \cite{ref8} into the phase diagram
of the ``dressed'' membrane. As is clear from comparison with the
general analysis \cite{ref7} the phase diagram of the polymer
embedding surface covers only one part of the phase diagram expected
on general grounds.

The tubulization transition, if indeed present in the systems with
strong adsorption of polymers to soft flexible surfaces, would be
of particular importance for understanding the DNA - cationic lipid
aggregation \cite{ref2, ref3}, where the colloidal state of the
aggregate is crucial for transfection of the DNA - lipid complex
across the cellular membrane.



\end{document}